\documentclass[12pt]{article} 

\usepackage{latexsym}
\usepackage[dvips]{graphicx}

\usepackage{stmaryrd}
\usepackage{amsmath}
\usepackage{amssymb, amscd}
\usepackage{bbold}
\usepackage{epsf}%
\usepackage{mathrsfs}

\newcommand{\be}{\begin{equation}}
\newcommand{\ee}{\end{equation}}
\def\bea{\begin{eqnarray}}
\def\eea{\end{eqnarray}}
\def\beq{\begin{eqnarray*}}
\def\eeq{\end{eqnarray*}}
\def\ba{\begin{array}}
\def\ea{\end{array}}

\renewcommand{\L}{{\mathscr L}}

\newcommand{\Res}{{\rm Res}}

\def\x{{\bf x}}

\def\K{{\bf K}}
\def\n{{\bf n}}
\def\P{{\bf P}}

\def\Lb{{\bf L}}

\def\bR{{\Bbb R}}
\def\bZ{{\Bbb Z}}

\def\tot{{\rm tot}}
\def\reg{{\rm reg}}
\def\ren{{\rm ren}}

\begin{document}

\begin{center}

{\bf DENSITY PERTURBATION IN THE UNIVERSE WITH
NONTRIVIAL TOPOLOGY OF SPACE-TIME}

\vspace{0.5cm}

S.S.~Moskaliuk, A.V.~Nesteruk and I.I.~Sokolov

\vspace{0.5cm}

{\it Bogolyubov Institute for Theoretical Physics\\
Metrologichna str. 14-b, 252143 Kiev 143, Ukraine\\
mss\@bitp.kiev.ua }

\end{center}

\begin{abstract}
 A space spectrum of density perturbation In the Universe with 
nontrivial topology of space-time is shown to become discrete.
\end{abstract}

\newpage

\section{Introduction}

The problem of cosmological density perturbations has a  long 
history and has been studied for a long time within the framework 
of both the Friedmannian cosmology and the inflationary one (see 
e.g. the reviews [1]). Since many cosmologists share now the 
point of view that the observed Friedmannian stage followed an 
inflationary one, the most important mechanisms of a generation 
of small density perturbations (necessary for galaxy formation) 
are those which work just at the inflationary stage. Much work 
has first been performed in studying the so-called adiabatic 
perturbations produced during an inflation. Then, however, it, 
was pointed out in Ref. [2] that it should be also important to 
explore the so-called isothermal perturbations generated during 
the inflation. In view of this, the authors of Ref. [3] have 
discussed several different mechanisms to generate both of 
perturbation types. The adiabatic perturbations are connected 
with the metric or curvature perturbations, while the isothermal 
ones (at constant energy density of system) turn out to be 
essential for the big scale factors of a model and can initiate 
the adiabatic perturbations.  Within the framework of an 
inflationary cosmology the adiabatic fluctuations of density p 
are usually related with the perturbations of a scalar field  
(the inflation field) responsible for the character of 
cosmological expansion, viz.

\beq
{\delta \rho\over \rho} \sim {\delta\Phi\over \Phi}.
\eeq

Within the realistic elementary particle theories there  exist 
the other scalar fields $\phi$ of different types. For example, 
the existence of the fields $\phi$, which very weakly interact with 
the other fields, is typical for the theories with axion field, 
for all models of particle physics based on $N = 1$ supergravity 
induced by superstrings, and so on. During inflation 
perturbations of all these fields will be generated as well. If 
we denote the density $\rho_\Phi$, $\rho_\phi$ which is referred, 
respectively, to the $\Phi$, $\phi$ fields, and 
$\rho_\tot$ is a total 
density, then the following supposition consists in that at first 
$\rho_\Phi\simeq \rho_\tot\gg\rho_\phi$
and the perturbations of $\phi$-fields will not give 
rise to substantial contribution in the adiabatic perturbations 
of $\rho_\tot$, and, as a consequence, in the metric 
(temperature $T$) perturbations. The perturbation of $\phi$-fields 
(isothermal perturbations) comes into play later. This is the 
result of the fact that energy density of nonrelativistic 
particles ($\sim T^3$) associated with the field $\phi$ decreases 
more slowly than energy density of photons ($\sim T^4$) as 
products of a decay of field $\Phi$. We shall note that a phase 
transition occurs only for field $\Phi$ owing to a large energy 
density of it. A study of isothermal perturbations can be 
important, in particular, with account taken of a large hidden 
mass of the Universe (e.g. of the inflation is an axion field 
which drives inflation [3,4]). One must have, therefore, some 
generation mechanisms of both type perturbations.

Let us now say that usually in investigating cosmological  
density perturbations one ignores a possible nontrivial topology 
of the Universe (cf.  Refs. [1-3]), to be more precise, it is 
tacitly supposed to be effective $\bR^4$, i.e. trivial. Ellis, 
however, as flat , spherical and hyperbolical Robertson-Walker 
Universes, but back as 1971 [6] (see also Ref. [7]) pointed out 
that the Robertson-Walker metrics (underlying both Friedmannian and
inflationary cosmologies) may be realized not only on topologies of the
form $\bR\times\Sigma$ ($\Sigma = \bR^3, S^3, H^3$) corresponding 
to, respectively, also on those of the form 
$\bR\times\Sigma/\Gamma$, where $\Gamma$ is a discrete group of 
isometrics for $\Sigma$.  Effects of nontrivial space-time 
topology could be essential at both the Friedmannian stage [8] 
and the inflationary one [9]. Let us consider the case of the 
$M_3 = \bR\times T^3$ topology, i.e. $\Sigma=\bR^3$, 
$\Gamma=\bZ^3$, $T=S^1$.  In particular, the Universe with 
$M_3$-topology might be generated as a result of a quantum 
creation in various models:  in ordinary (nonsupersymmetric) 
field theories [10] i in different supergravity versions [11], 
in superstring theory [12], in supermembrane theory [13].

It is important that the nontrivial $M_3$-topology involves  the 
appearance of topologically inequivalent configurations (TICs) of 
real scalar fields [14]. The number of such TICs is equal to the 
number of elements of $H^1(M_3,\bZ_2)$, the first cosmology group 
of $M_3$ with coefficients in $\bZ_2$, and since 
$H^1(M_3,\bZ_2)=\bZ_2^3$, then that number is eight. There 
exist, therefore, 8 types of real scalar fields in $M_3$ (7 of 
which are twisted).

The following is now noteworthy. Since a real or complex line bundle
may have a cross-section that vanishes nowhere only in the case when the
line bundle is trivial [see e.g. Refs. |14,15]), hence it  
follows that twisted real scalar fields (corresponding to the 
cross-sections of nontrivial line bundles) must vanish for at 
least one point in $M_3$, and, as a consequence, the condition 
$\varphi_i = {\rm const}\neq 0$ can never be satisfied in the 
entire $M_3$.  In other words, any scalar twisted field 
$\varphi_i$ may be homogeneous in $M_3$ only with the value 
$\varphi_i=0$ due to topological reasons.  Only the field 
$\varphi_0$ (untwisted) has no limitations in a region of 
changing its values and may be constant in the whole $M_3$ which 
is needed for a realization of the chaotic inflation [5].

In the paper [16] the first attempt to study possible consequences
of bringing the aspect of nontrivial topology to the problem of density
perturbations was undertaken. There was considered an evolution of the
early Universe with $M_3$-topology and metric
\begin{equation}
ds^2=dt^2-a^2(t)\sum_{j=1}^3\, dx_i^2,
\end{equation}
and filled with a real scalar field $\varphi(x)$. Owing to the 
above observations the words "real scalar field" should actually 
stand for all the TICs of real scalar field admissible in the 
given topology and, as a result, practically stand for the set of 
$\varphi_i\, (i=0,1,\ldots,7)$ fields permissible by the 
$M_3$-topology (see above).  The authors naturally came to the 
expression for potential of theory, and, thus, the number of 
fields $\varphi_i$ was defined by space-time topology and without 
 introduction ad hoc.  The relation between interaction constants 
of the fields $\varphi_i$, and the classification of perturbations (both 
adiabatic and isothermal ones) were also naturally derived from a 
single foundation-stone, i.e. the nontrivial topology of 
space-time.

It was also assumed there that the scales $L_i\, (i=1,2,3)$ 
of the nontrivial topological structure will be much larger than 
modern horizon (an observable size of the Universe at present 
time) and $L_i$ are increasing as $a(t)$ in time. That is why the 
temporal evolution of fields and density perturbations will be 
just the same, as in the case of trivial topology.  This fact 
gives us a key to trace the second essential change in the 
theory, namely, that a space spectrum of density perturbations 
becomes discrete and dependent on $L_i$ --- it is a subject of 
considerations in this paper.

\section{The Amplitude and Spectrum of Density Perturbations  in 
Space-Time with Nontrivial Topology}

Let us consider noninteracting scalar field in space-time with 
$M_3$-topology of space in metric (1) with the scale factor 
$a(t)\sim e^{Ht}$ and Lagrangian


\beq
\L =\sum_{i=0}^7 \left[ {1\over2} \partial_\mu\varphi_i(x) 
\partial^\mu\varphi_i(x)- V_i\left(\varphi_i(x)\right)\right].
\eeq
In accordance with the above mentioned ideas we have 
really 8 types of topologically inequivalent configurations of 
scalar field.  Every field can be presented as a sum of classical 
part $\varphi_c$ and small quantum perturbation
\beq
\varphi_i(x)=\varphi_{ic}(x) +\varphi_{iq}(x) \qquad 
i=0,1,\ldots, 7.
\eeq
It  is known (see, for example, [3]), that after inflation the 
classical part $\varphi_c$ becomes spatially homogeneous 
$\varphi_{ic}(x)=\varphi_{ic}(t)$. Because of the above 
properties of TICs of real scalars, only untwisted configuration 
$\varphi_0$ may •take any values constant in the whole $M_3$ 
while the fields $\varphi_i\, (i\neq0)$ may be constant only with 
values $\varphi_i = 0$ in the whole $M_3$.  It means that after
inflation

\beq
\varphi_{ic}=0, \qquad \qquad i=1,\ldots,7.
\eeq
The  field $\varphi_{0C}=\varphi_{0C}(t)$ identified with an 
inflation and it satisfies equation

\beq
\ddot{\varphi}_{0C}+3H\dot{\varphi}_{C1} =-{dV_0(\varphi_0)\over 
d\varphi_0}\, ,
\eeq
whereas small perturbations $\varphi_{gi}$ of this field satisfy 
equation

\beq
\ddot{\varphi}_{qi}+3H\dot{\varphi}_{gi} 
-e^{2Ht}\nabla^2\varphi_{gi}
 &\!\!\!=\!\!\!& -\left( {d^2V_i\over d\varphi_i^2}\right) 
\varphi_{gi} \equiv -m^2\left(\varphi_{ci}\right) \varphi_{qi}\\
i &\!\!\!=\!\!\!& 0,1,\ldots,7.
\eeq

To separate variable in this equation we present $\varphi_{qi}$ 
in the form

\beq
\varphi_{qi}(x) =\psi_i\left(t,K^{(i)}\right) \exp 
\left(i\K^{(i)}\x\right).
\eeq
Then for time-dependent part of the field $\varphi_{qi}$ 
subjected to normalization condition

\begin{equation}
i\int\limits_0^\Lb\, \sqrt{-g}\, 
d^3x\psi_i^*\Bigl(t,K(i)\Bigr)
\overleftrightarrow{\partial t}\psi_i\Bigl(t,K'_{(i')}\Bigr) 
=\delta K^{(i)}K^{'(i)} \delta_{ii'},
\end{equation}
one can obtain [17]
\begin{equation}
\psi_i\Bigl(t, K^{(i)}\Bigr) ={H|\eta|^{3/2}\over 
\sqrt{L_1L_2L_3}} \, {\sqrt{\pi}\over2} \left[ C_{11} H_\nu^{(1)} 
\left(K^{(i)}\eta\right) +C_{2i}H_\nu^{(2)} 
\left(K^{(i)}\eta\right)\right],
\end{equation}
where $\eta=-H^{-1}e^{-Ht}$, $\nu^2={9\over4}-{m^2\over H^2}$, 
$H_\nu^{(1)}$ and $H_\nu^{(2)}$ are Hankel functions.  $H$ is the 
Bubble parameter.  Later oil we shall consider only the case when 
$m^2\ll H^2$ and that is why $\nu = 3/2$ and normalization 
condition (2) will
look like
\beq
|C_{2i}|^2 -|C_{1i}|^2=1.
\eeq

To define a spectrum $K^{(i)}$ we have take into account the  
transformation properties of TICs:
\beq
\varphi_{qi}\left(t,x_1+L_1n_1, x_2+L_2n_2, x_3+L_3n_3\right)= 
(-1)^{\lambda(i)} \varphi_{qi}\left(t,x_1,x_2,x_3\right),
\eeq
\beq
\ba{llll}
\lambda^{(0)} = 0,& \lambda^{(1)}=n_1, & \lambda^{(2)}=n_2,& 
\lambda^{(3)}=n_3,\\[0.4cm]
\lambda^{(4)}=n_1+n_2,& \lambda^{(5)}=n_1+n_3,& 
\lambda^{(6)}=n_2+n_3,&\\[0.4cm]
\lambda^{(7)}=n_1+n_2+n_3,&& n_j=0,\pm1,\pm2,\ldots . &\\
\ea 
\eeq
Hence, one can deduce the formula for the spectrum $K^{(i)}$
\beq
\ba{lll}
K_j(i)={2\pi\over L_j} \Bigl( n_j+g_j^{(i)}\Bigr), && n_i\in Z,\, 
\quad j=1,2,3,\\[0.4cm]
g_j^{(0)}=0, & g_i^{(l)}=(1/2)\delta_{j\iota},& l=1,2,3, \,  
j=1,2,3,\\[0.4cm]
g_j^{(4)}=(1/2)(\delta_{1j} +\delta_{2j}), & 
g_j^{(5)}=(1/2)(\delta_{1j} +\delta_{3j}), &\\[0.4cm]
g_j^{(6)}=(1/2)(\delta_{2j} +\delta_{3j}), & g_j^{(7)} =(1/2) 
(\delta_{1j}+\delta_{2j} +\delta_{3j}). &\\ \ea
\eeq
Now we are to calculate the amplitude of the density 
perturbations of TICs
\beq
\delta\varphi_{qi}^2 =\langle 0|\hat{\varphi}^2_{qi}|0\rangle,
\eeq
where $\hat{\varphi}_{qi}$ is the corresponding operator of field 
after the second quantization, $|0\rangle$ is the de Sitter 
invariant vacuum where $C_{1i}$, and $C_{2i}$ in formula (3) are 
chosen to be 0 and 1, respectively, (see also [17]).

By analogy with [17,18] the value of $\delta\varphi_{qi}^2$ one 
can calculate
\begin{equation}
\delta\varphi_{qi}^2(t) =\sum'_{n_1,n_2,n_3\in Z}\, |\psi_i\Bigl( 
t,\K^{(i)}\left( \{n_j\}^3_{j=1}\right)\Bigr)|^2\, ,
\end{equation}
where the prime signifies that term with $n_j = 0$ when 
$g_j^{(i)} = 0$ should be omitted. Substituting here the 
expression (3) and writing (4) in terms of physical momentum 
$P_\n=e^{-Ht}K_\n$ and topological radius $L(t) = e^{Ht}L$ we 
come to the expression for the amplitude of density 
perturbations $(L_1=L_2=L_3)$

\begin{eqnarray}
\left(\delta\varphi_{qi}\right)^2  &\!\!\!=\!\!\!& \Delta_0^{(i)} 
+\Delta_1^{(i)},\nonumber \\
\Delta_0^{(i)}  &\!\!\!=\!\!\!&  
{1\over 2l^3(t)}\, \sum'_{\n\in 
Z}\, {1\over P_n^{(i)}} =\nonumber \\
 &\!\!\!=\!\!\!& {1\over (2\pi) 2L^2(t)}\, \sum'_{\n\in Z}\,
\Biggl[\left( n_1+g_1^{(i)}\right)^2 
+\left(n_2+g_2^{(i)}\right)^2 +\left(n_3+g_3^{(i)}\right)2 
\Biggr]^{1/2}, \nonumber \\ 
\Delta_1^{(i)}  &\!\!\!=\!\!\!& 
{H^2\over 2L^3(t)}\, \sum'_{\n\in Z}\, {1\over (\P_n^{(i)})^3} 
=\nonumber \\
 &\!\!\!=\!\!\!& {H^2\over 2(2\pi)^3}\, \sum'_{\n\in Z}\, 
\Biggl[\left(n_1+g_1^{(i)}\right) ^2 
+\left(n_2+g_2^{(i)}\right)^2 
+\left(n_3+g_3^{(i)}\right)^2\Biggr]^{-3/2}, \nonumber \\ 
P_\n^{(i)}  &\!\!\!=\!\!\!& {2\pi\over L(t)} \Biggl[ \sum_{j=1}^3 
\left(n_j+g_j^{(i)}\right)^2\Biggr]^{1/2}.
\end{eqnarray}


It is easily seen, that when $H = 0$, i.e. we have a usual 
Minkovski space, then $\Delta_1^{(i)}=0$ and 
$\Delta_0^{(i)}\neq0$, $\Delta_0$ represents an oscillating vacuum
contribution from oscillating vacuum quantum fluctuations. It is also
easy to understand that the right-hand expressions for 
$\Delta_0^{(i)}$ and $\Delta_1^{(i)}$ are
divergent. It means, that we have to find a way to interpret these formulae
correctly. This procedure will be some kind of regularization of
$\Delta_0^{(i)}$ and
$\Delta_1^{(i)}$ and  will be found on the formalism of Epstein 
zeta functions.

We define the 3-dimensional Epstein zeta function for $\Res>1$ by 
the formula
\begin{equation}
Z_3\Biggl| \ba{ccc} g_1& g_2& g_3\\ 0&0&0\\ \ea \Biggr|(s) 
=\sum'_{\n\in Z} \Biggl[ \left(n_1+g_1\right)^2 
+\left(n_2+g_2\right)^2 +\left(n_3+g_3\right)^2 
\Biggr]^{-{3S\over2}}.
\end{equation}
For $\Res<1$, $Z_3(s)$ is understood as the analytical 
continuation of the right-hand side of (6). $Z_3(s)$ satisfies 
the functional equation
\begin{eqnarray}
 &\!\!\!\!\!\!&  
\Gamma \left({3\over2} s\right) Z_3 \Biggl|\ba{ccc}
g_1& g_2& g_3\\ 0&0&0\\ \ea\Biggr|(s) =\pi^{3s-3/2} 
\Gamma\left({3\over2}(1-s)\right)\nonumber \\
 &\!\!\!\times\!\!\!&  Z_3\Biggl| \ba{ccc}
0&0&0\\ -g_1& -g_2& -g_3\\ \ea \Biggr|(1-s).
\end{eqnarray}
Formally we can write the expressions for $\Delta_0^{(i)}$ and 
$\Delta_1^{(i)}$ in terms of 
$Z_3(s)$, where $s=1/3$ for $\Delta_0^{(i)}$ and $s=1$ for 
$\Delta_1^{(i)}$. 
($Z_3(s=1)$ has a simple pole.)  Taking equation (7) we in fact 
regularize $\Delta_0^{(i)}$ and $\Delta_1^{(i)}$

\begin{eqnarray}
\Delta_{0\reg}^{(i)}  &\!\!\!=\!\!\!& {1\over(2\pi)^2L^2(t)} 
Z_3\Biggl|\ba{ccc} 0&0&0\\ -g_1^{(i)} & -g_2^{(i)} & -g_3^{(i)}\\ 
\ea \Biggr|\left({2\over3}\right),\nonumber \\[0.4cm]
\Delta_{1\reg}^{(i)}(s) &\!\!\!=\!\!\!& 
{H^2\over2(2\pi)^2} Z_3\Biggl|\ba{ccc} 0&0&0\\
-g_1^{(i)}& -g_2^{(i)}& -g_3^{(i)}\\ \ea\Biggr|(s)\Gamma(s). 
\end{eqnarray}

The value of $\Delta_0^{(i)}$ is finite and equals 
$\alpha_i/(2\pi)^2L^2(t)$ for different $i$, where (according to 
[15])
\beq
\alpha_0 &\!\!\!=\!\!\!& -8.91363\ldots \, ,\\
\alpha_1 &\!\!\!=\!\!\!& \alpha_2=\alpha_3=-0.301380\ldots\, ,\\
\alpha_4 &\!\!\!=\!\!\!& \alpha_5=\alpha_6=-1.83004\ldots \, ,\\
\alpha_7 &\!\!\!=\!\!\!& -2.51935\ldots \, .
\eeq

To transform the expression (8) further one should use the 
presentation  of $\Gamma(s)$ in the vicinity of $s=0$

\beq
\Gamma(s)\simeq {1\over s}-C+0(s), \qquad C= 0.57721\ldots \,.
\eeq
We can rewrite
\begin{equation}
\Delta_{1\reg}^{(i)} =\beta_i{H^2\over 2(2\pi)^2} 
\left({1\over s}-C\right) +0(s),
\end{equation}   
where
\beq
\beta_i=Z_3\Biggl| \ba{ccc} 0&0&0\\[0.3cm]
-g_1^{(i)}& -g_2^{(i)}& -g_3^{(i)}\\ \ea \Biggr| (0).
\eeq
The infinity in (9) is removed by the Hubble constant $H$ 
renormalization

\beq
H^2\left[ 1-C^{-1}s^{-1}\right] \rightarrow J_\ren^2\, ,
\eeq
where $H_\ren$ is the physical value of Hubble parameter. 
Finally

\beq
\Delta_{1\ren}^{(i)} ={H^2_\ren\over 2(2\pi)^2}\, \beta_i C.
\eeq

\section{Conclusion}

Considering the final values of $\Delta_{0\ren}^{(i)}$ and 
$\Delta_{1\ren}^{(i)}$
\beq
\Delta_{0\ren}^{(i)}  &\!\!\!=\!\!\!& {\alpha_i\over 
(2\pi)^2L^2(t)},\\
\Delta_{1\ren}^{(i)}  &\!\!\!=\!\!\!& {H_\ren^2\over 2(2\pi)^2}\, 
\beta_i C
\eeq
we can conclude that contribution from $\Delta_{0\ren}^{(i)}$ to 
$\delta\varphi_{qi}^2$ after inflation is
exponentially small, whereas the value of $\Delta_{1\ren}^{(i)}$ 
is constant and, as it was in space-time with trivial topology, 
is proportional to squared  Hubble parameter [3,17,18]. The 
result for $\delta\varphi_{qi}^2$ is derived just without any 
additional suppositions, and it is exact one. We see so that 
inclusion of nontrivial space-time topology does not lead to 
principally new results:  the amplitude of fluctuations is the 
same up to the finite numerical  coefficient. The only difference 
is connected with the discrete character of the spectrum of 
perturbations which is given by formula (5).

The authors are grateful to Yu.P.~Goncharov for helpful 
discussions.

\end{document}